\documentclass[12pt]{article}
\usepackage{graphics}
\renewcommand{\theequation}{\arabic{section}.\arabic{equation}}
\begin{document}
\markright{Hidden variables or...}
\title{Hidden Variables or Positive Probabilities?}
\author{Tony Rothman~$^*$ and E.\ C.\ G. Sudarshan~$^\dagger$\\[2mm]
 {\small \it
\thanks{trothman@titan.iwu.edu}~
~Dept.\ of Physics, Illinois Wesleyan University, Bloomington, IL
61702, USA.}\\ {\small
\it
\thanks{sudarshan@physics.utexas.edu}~Dept.\ of Physics, University of
Texas, Austin TX 78712, USA}\\
 }
\date{{\small   \LaTeX-ed \today}}

\maketitle

\begin{abstract}
Despite claims that Bell's inequalities are based on the Einstein
locality condition, or equivalent, all derivations make an
identical mathematical assumption: that local hidden-variable
theories produce a set of positive-definite probabilities for
detecting a particle with a given spin orientation.  The standard
argument is that because quantum mechanics assumes that particles
are emitted in a superposition of states the theory cannot produce
such a set of probabilities.  We examine a paper by Eberhard, and
several similar papers, which claim to show that a generalized
Bell inequality, the CHSH inequality, can be derived solely on the
basis of the locality condition, without recourse to hidden
variables.  We point out that these authors nonetheless assumes a
set of positive-definite probabilities, which supports the claim
that hidden variables or ``locality" is not at issue here,
positive-definite probabilities are. We demonstrate that quantum
mechanics does predict a set of probabilities that violate the
CHSH inequality; however these probabilities are not
positive-definite. Nevertheless, they are physically meaningful in
that they give the usual quantum-mechanical predictions in
physical situations. We discuss in what sense our results are
related to the Wigner distribution.

\vspace*{5mm} \noindent PACS: 03.65-w,03.65.Bz
\end{abstract}

\section{Introduction}
\label{sec1}

With the introduction of his celebrated inequalities in 1964, John
Bell \cite{Bell64} provided the basis for an experimental test to
distinguish quantum mechanics from local hidden-variable theories.
Since that time the universal interpretation of the results has
been that quantum mechanics violates Bell's inequalities due to
its ``nonlocal" character, whereas local hidden variable theories
satisfy the inequalities because, as their name implies, they are
``local."

The situation is actually not so transparent. Bohr taught us to be
aware of ambiguous language. Although derivations of Bell's
inequalities are evidently based on Einstein's ``locality"
condition, couched in various phrases such as ``principle of
separability" and so forth, mathematically all derivations make
an identical assumption, specifically: {\em hidden-variable
theories introduce a set of {\rm a priori} positive-definite
probabilities P that are not predicted by quantum mechanics.} In
Bohm's classic version of the Einstein-Podolsky-Rosen experiment,
for example, a particle in a spin-singlet state decays into two
daughter particles with zero total angular momentum (see, e.g.,
Sakurai's text \cite{Sakurai} or Sudarshan and Rothman
\cite{SR93}, henceforth SR). According to local hidden-variable
theories there is an {\it a priori} positive-definite probability
that the daughter particles will be detected with spins ``up"
along a chosen axis. Quantum mechanics, on the other hand,
assumes that the daughter particles are in a superposition of
states and so, by definition, there can be no {\it a priori}
probability P such that their spins will be detected along a
given direction.

Contrary to this view, in SR we pointed out that quantum mechanics
{\em does} predict a set of {\it a priori} probabilities, in
exactly the same way as do hidden-variable theories, but the
quantum probabilities are not positive-definite.  They are
nevertheless meaningful in that when applied to physical
situations they give the standard quantum-mechanical answers, in
particular the usual violation of Bell's inequalities. Given the
exact analogy in producing the two sets of probabilities the
distinction between ``local" hidden-variable theories and
``non-local" quantum mechanics is dissolved.  From this point of
view one merely has two competing theories that give two
different sets of probabilities; it is unsurprising that
hidden-variables theories fail experimental tests of Bell's
inequalities because they used the wrong set of probabilities for
a quantum-mechanical problem.

The notion of ``extended" probabilities dates back to Dirac and we
have not been the only authors to suggest that they can resolve
the EPR paradox (see \cite{Muck82,Muck86})
 but, needless to say, the SR argument has
not found widespread acceptance. Recently, several rather old
papers, in particular one by Eberhard \cite{Eberhard77}
 entitled ``Bell's Theorem
Without Hidden Variables," have come to our attention. Eberhard's
paper is of interest because it claims to show that a more
general version of Bell's inequalities, known as the CHSH
inequality (after Clauser, Horne, Shimony and Holt) \cite{CHSH69},
is violated by quantum mechanics, and that the CHSH inequality can
be demonstrated solely on the basis of the locality principle,
{\em without the introduction of hidden variables}.  (A slightly
later paper by Peres \cite{Peres78} gives an almost identical
argument; one by Stapp \cite{Stapp85} is in some respects
similar.) At first sight these proofs appear to assume little more
than $2 < 2\sqrt 2$. On closer inspection, however, we find that
they ``play into our hands," i.e., they may not make an explicit
statement about hidden variables but they {\em do} assume a set of
positive-definite probabilities. We now demonstrate this is so,
reinforcing the contention in SR that, despite any words employed,
the crucial {\em mathematical} assumption in derivations of Bell's
inequalities is not locality but positive probability.

\section{The Eberhard Argument}
\label{sec2}

Eberhard considers two identical apparata, $A$ and $B$, at two
different locations.  On apparatus $A$ is a knob $a$ that can be
turned to two positions, 1 and 2.   On apparatus $B$ is a knob $b$
that can also be turned to two positions, 1 and 2.  With its knob
at either position apparatus $A$ can record a series of events. It
is not important exactly what the events are, but we assume that
for each event each apparatus can measure only one of two possible
outcomes, which for simplicity we take to be $\pm 1$. When the
knob $a$ is in the 1 position, we designate the outcome of the
$jth$ event as $\alpha_{1j}$, with similar notation for position 2
and knob $b$. For each event we can thus in principle have:
$\alpha_{1j} = \pm 1,\ \alpha_{2j} = \pm 1,\ \beta_{1j} = \pm 1,
\beta_{2j} = \pm 1$. However, for each measurement we will choose
only one setting on each apparatus, so a given event will produce
a pair of readings, such as $\alpha_1 = 1, \beta_2 = -1$. (Here
and below we suppress the subscript $j$ when it will not cause
confusion.)

For a series of $N$ measurements Eberhard next defines a quantity
$C$, such that
\begin{equation}
C = \frac{1}{N}\sum_{j=1}^N\alpha_j\beta_j
\end{equation}

We see that $C = <\alpha_j\beta_j>$, the statistical mean of the
$N$ products $\alpha_j\beta_j$.  No restriction is placed on the
fraction of the $N$ measurements for which the $\alpha$'s and
$\beta$'s come out positive or negative, but note that each
product $\alpha_j\beta_j = 1$ when $\alpha$ and $\beta$ have the
same sign and $\alpha_j\beta_j = -1$ when they have opposite
signs. Thus $C$ represents the fraction of events in which
$\alpha$ and $\beta$ have the same sign minus the fraction in
which they have opposite sign.

Because each knob has two positions, there are four possible
versions of $C$.  That is, we can define
\begin{eqnarray}
 C_{11} & = & <\alpha_{1}\beta_{1}> \nonumber \\
C_{12} & = & <\alpha_{1}\beta_{2}>  \nonumber \\
 C_{21} & = & <\alpha_{2}\beta_{1}>  \nonumber \\
 C_{22} &  = & <\alpha_{2}\beta_{2}>
  \label{C}
 \end{eqnarray}
(sum on $j$ understood). Here, $C_{11}$ is  just the above
statistical mean when knobs $a$ and $b$ are both in position 1,
and so forth.

Now, for each event let
\begin{equation}
\gamma \equiv \alpha_1\beta_1 + \alpha_1\beta_2 +
                \alpha_2\beta_1 - \alpha_2\beta_2 .
\label{gamma}
\end{equation}
Then, the statistical mean of $\gamma$ is just
\begin{eqnarray}
<\gamma> & = & \frac{1}{N}\sum_{j = 1}^N \gamma_j \nonumber \\
            & = & \frac{1}{N}\sum_{j = 1}^N
                     (\alpha_1\beta_1 + \alpha_1\beta_2 +
                \alpha_2\beta_1 - \alpha_2\beta_2) \nonumber \\
            & \equiv & C_{11} + C_{12} + C_{21} - C_{22},
            \label{gammaavg}
\end{eqnarray}
where in the second line we have again suppressed $j$.

 The locality condition enters the discussion when we attempt to put
bounds on $<\gamma>$. Recall that a knob will be set to either
position 1 or 2 for each measurement. We assume that a measurement
on $A$ is independent of a measurement on $B$. The $\alpha$'s and
$\beta$'s are thus treated independently. This is the locality
condition.

At this point a digression is necessary. Eberhard states that only
one setting of each knob (position 1 or 2) will be used for each
measurement, and that thus only one $\alpha$ or $\beta$ is
recorded for each event. However, if this were indeed the case,
then for each measurement only one term in $\gamma$ would survive
(one product $\alpha\beta$) and the upper bound on $\gamma$ would
be 1 (cf. Eqs. (\ref{gamma}) and (\ref{triangle})). That the upper
bound is 2 shows that {\em mathematically} all four possible terms
$\alpha\beta$ are present in $\gamma$. Consequently, not only are
the $\alpha$'s being taken to be independent of the $\beta$'s but
$\alpha_1$ ($\beta_1$) is being treated as independent of
$\alpha_2$ ($\beta_2$). The rationale for including all $\alpha$'s
and $\beta$'s in $\gamma$ simultaneously comes from a 1971
suggestion of Stapp \cite{Stapp71}. Stapp, Eberhard (and Peres
\cite{Peres78} in his nearly identical thought experiment), are
actually considering all possible outcomes of the measurements in
a hypothetical ensemble space.  By doing so they intend to show
that any conceivable outcome of the experiment is violated by
quantum mechanics.

One can take several attitudes toward such a procedure. A first
possible attitude is that it is illegitimate to speculate about
the results of unperformed experiments.  In other words, if one
takes the quantity $\gamma$ literally, the knobs must be set to
two positions at once, a physical impossibility.  A second view is
that it is indeed legitimate to think about all possible outcomes
of an experiment\footnote{This concept is often referred to as
``counterfactual definiteness," after Stapp.} and that if one does
so, one is forced to the conclusion that quantum mechanics is
nonlocal. In fact, there is a third possible viewpoint.  As we
discuss below, the $\gamma$'s are derivable from the ``master
probabilities" employed in a standard derivation of Bell's
inequalities, quantities that are not directly measurable but
nevertheless have physical consequences. Hence both the Eberhard
procedure and the standard derivation suffer from exactly the same
ambiguities. For the moment it is not important which philosophy
one adopts; we merely treat $\gamma$ as a mathematical quantity,
as Eberhard does. At the same time, however, we see that by
treating {\em all} the $\alpha$'s and $\beta$'s as independent,
mathematically the locality condition becomes indistinguishable
from the general assumption of independent variables.

In any case, following Eberhard we assume 16 possible values for
each $\gamma$. At this stage of the exposition, Eberhard goes
through an elaborate argument to show that $\gamma \le 2$ always.
However, let us redistribute the terms in Eq. (\ref{gamma}) and
write
\begin{equation}
\gamma = \alpha_1(\beta_1 + \beta_2) + \alpha_2(\beta_1 -
                                                        \beta_2).
\label{gambound}
\end{equation}
Because $\beta_1$ and $\beta_2$ are equal or of opposite sign, if
the first term is nonzero, the second term is zero and vice versa.
Thus we can see trivially that $\gamma = \pm 2$ always and
$|\gamma | = 2$, period.

But by the triangle inequality we know that
\begin{equation}
|\frac{1}{N}\sum_{j=1}^N (\alpha_1\beta_1 + \alpha_1\beta_2
+\alpha_2\beta_1 - \alpha_2\beta_2)|
 \le \frac{1}{N}\sum_{j=1}^N |(\alpha_1\beta_1 + \alpha_1\beta_2
+\alpha_2\beta_1 - \alpha_2\beta_2)|
\end{equation}
Yet from Eq. (\ref{gammaavg}) and Eq. (\ref{gamma}) this is by
definition
\begin{eqnarray}
|C_{11} + C_{12} + C_{21} - C_{22}| &\le & \frac{1}{N}\sum_{j=1}^N
                                            |\gamma_j|  \nonumber \\
                                    & = & \frac{1}{N} \times N
                                    \times 2
                                    \label{triangle}
\end{eqnarray}
The CHSH inequality follows immediately:
\begin{equation}
|C_{11} + C_{12} + C_{21} - C_{22}| \le 2 , \label{CHSH}
\end{equation}
or, in more compact notation,
\begin{equation}
|{\cal C}| \le 2 .
 \label{CHSHb}
\end{equation}

Eberhard next considers a quantum-mechanical experiment in which
two photons are emitted in the directions of $A$ and $B$ by an
atom between them. The photons are detected by polarizers; each
$\alpha$ ($\beta$) is taken to be +1 when one polarization is
detected and -1 when the other is detected. Unfortunately, at this
point the paper becomes very unclear. Eberhard merely asserts
without calculation that for each of the $C$'s in Eq. (\ref{C}),
quantum mechanics predicts that ``if the number of events $N$ is
large enough, then $C \cong cos(2a - 2b)$," where $2a-2b$ is twice
the angle between the polarizers. Actually, no approximation is
necessary. For spin-1/2 particles, the correct prediction is
\begin{equation}
{\cal C}_{qm} =  3 cos\theta - cos3\theta,
                                    \label{Cqm}
\end{equation}
which we derive below, and in which $\theta$ is the angle between
polarizers. (The result for photons will be the same if $\theta$
is taken to be twice the angle between polarizers.) Note that for
$\theta = 45^o$ (\ref{Cqm}) gives ${\cal C}_{qm} = 2\sqrt 2 \ge
2$. Therefore, quantum mechanics violates the CHSH inequality,
just as it does the Bell inequalities.

 As mentioned above, the demonstration seems to assume almost nothing:
no hidden variables, merely ``locality," which implies that a
certain mathematical quantity $\gamma$ always equals $\pm$ 2.
However, on closer inspection we find that more than an assumption
of independent $\alpha's$ and $\beta$'s is being made. In the
first place, the value 2 on the right-hand side of Eq.
(\ref{CHSH}) is entirely arbitrary and results merely from the
choice of $\pm 1$ as the ``eigenvalues" for $\alpha$ and $\beta$.
One could have equally well chosen $\pm 1000$. In that case,
however, one would necessarily have to assume that the
corresponding quantum experiment also had eigenvalues of $\pm
1000$.  This matter is not so serious, but it nevertheless
illustrates that the CHSH inequality is not a purely mathematical
assertion; a real measurement does lurk in the background.

The central issue lies elsewhere.  Eberhard's version of CHSH
inequality is a statement about the statistical mean of $\gamma$,
and therefore it {\em does} deal with a probability distribution
over the $\gamma$. Moreover, the frequency that a particular
$\gamma$ occurs is clearly taken to be positive. That
probabilities should be positive-definite is usually regarded as
self-evident, but because the assumption is the crux of the
matter, we spend a moment examining it. (In the Appendix we detail
where other authors have made the same assumption.)

As mentioned, there are 16 possible combinations of
$\alpha_1\beta_1 + \alpha_1\beta_2 +\alpha_2\beta_1 -
\alpha_2\beta_2 \ (= \gamma)$, of which eight have the value +2
and eight have the value -2. In a sequence of $N$ measurements,
let us suppose that +2 occurs $n_1$ times and -2 occurs $n_2$
times, such that $n_1 + n_2 = N$. Then
\begin{equation}
{\cal C} = \frac{2}{N}[n_1 -n_2].
\end{equation}
If all frequencies are equal, i.e. $n_1 = n_2$, then ${\cal C} =
0$. If $n_2 = 0$, then ${\cal C} = 2$ and if $n_1 = 0$ then ${\cal
C} = -2$. But here we have assumed that both $n_1$ and $n_2$ are
positive-definite. If $n_2 < 0$, then ${\cal C} > 2$. In other
words, the step leading to the second line in Eq. (\ref{triangle})
is valid only when $|n| = n$.

The notion of ``extended" (non-positive-definite) probabilities
has been considered by a surprising number of prominent
investigators, but the majority of physicists continue to regard
them with distaste, if not revulsion. Nevertheless, the quantum
violation of the bound on $\cal C$ is effectively due to the fact
that quantum mechanics allows negative probabilities. In the next
section we examine this claim in greater detail.

\section{Quantum Mechanical Probabilities}
\setcounter{equation}{0}
 \label{QMP}

Before deriving Eq.(\ref{Cqm}), it will be helpful to summarize
the procedure for obtaining the standard Bell Inequalities in
order to point out similarities to the CHSH-Eberhard experiment.
The reader is referred to SR or Sakurai \cite{Sakurai} for
additional details; see also the Appendix. Like its successor,
Bell's theorem is valid for local hidden-variable theories, which
involve only classical probabilities. In a typical derivation
such as Sakurai's one assumes that spin measurements may be made
along any of three axes, {\bf a}, {\bf b} and {\bf c}. A system
of decaying atoms emits $N$ particles of which a certain fraction
are taken to be, say, of the type ({\bf a}+, {\bf b}+, {\bf c}+)
$\equiv (+++)$, which designates spin up along all three axes. To
ensure zero total angular momentum, each emitted particle of type
(+++) must be paired with one of type ($ - - - $). There are
eight such spin combinations in all, as listed in Table 1.

The probability that (+++) is emitted (and in the case of hidden
variables, detected) is defined simply as $P(+++)= N(+++)/N$. One
can immediately object that such a probability is unphysical
because to determine it requires three simultaneous spin
measurements on a system of two particles, which is impossible. To
eliminate this difficulty, one forms pairwise probabilities of the
type $P({\bf a}+,{\bf b}+)\equiv P(++)$, which represents the
joint probability that the first particle will be found + along
{\bf a} and the second particle + along {\bf b}. This is easily
done. From the table, the total number of particles such that the
first particle's spin is + along {\bf a} is $N(+-+) + N(+--)$,
which must be paired with $N(-+-) + N(-++)$, the total number of
particles for which the second particle's spin is  + along {\bf
b}. This combination is labeled $N_3 + N_5$. Next one forms
triangle-type inequalities such as
\begin{equation}
N_3 + N_5 \le (N_2 + N_5) + (N_3 + N_7), \label{triangleN}
\end{equation}
which is obviously true, since we have just added positive numbers
to $N_3 + N_5$.  Dividing by $N$ gives by definition
\begin{equation}
  P({\bf a}+,{\bf b}+) \le P({\bf a}+,{\bf c}+) + P({\bf c}+,{\bf b}+),
  \label{Bell}
\end{equation}
one of the Bell inequalities. Eq. (\ref{Bell}) involves only one
measurement on each particle and so represents a physically
realizable situation. Note that the ``three-probabilities"
$P(+++)$ were reduced to pairwise probabilities $P(++)$ by summing
over the spins on the extraneous axis, in the above example {\bf
c}. We emphasize that, just as was the case for the CHSH
inequality, the Bell inequality is valid only if the N's and hence
the P's are taken to be positive-definite. In SR we demonstrated
that one can form {\em quantum probabilities} $P(+++)$, analogous
to the classical probabilities, then sum over the third argument
exactly as above to get pairwise quantum probabilities $P(++)$
that violate (\ref{Bell}) in the usual way.

By this point the reader will have noticed a similarity between
the $\gamma$'s in Eberhard's experiment and the
three-probabilities here.  Authors who derive the generalized Bell
inequalities introduce $\gamma$ as a measure of correlations
between real and imagined experiments but, as mentioned, if one
takes it literally it amounts to having the apparatus knobs set on
two positions simultaneously. This would seem to represent the
same sort of physical impossibility as that of making three
simultaneous spin measurements on two particles. Indeed, we will
demonstrate in Section \ref{Conclusions} that the two procedures
are identical: Introducing an ensemble of hypothetical
measurements is exactly equivalent to assuming a ``master
probability distribution" that requires more than two simultaneous
spin measurements on two particles. Before doing so, however, we
return to the Eberhard derivation.

Eberhard's experiment involves four axes, ${\bf a_1, a_2, b_1,
b_2}$, rather than three, but otherwise is almost identical to
the standard derivation of Bell's inequalities and so it is not
surprising that the above procedure can be followed to
demonstrate a violation of the CHSH inequality. We first need to
compute the quantum pairwise probabilities of the type just
mentioned, $P({\bf a}+,{\bf b}+)$. There are several ways to do
this. Following SR, we write the quantum-mechanical projection
operator for spin-1/2 particles as
\begin{equation}
\Pi({\bf a}\pm) = \frac{1}{2}({\bf 1} \pm \mbox{\boldmath
                                        $\sigma\cdot a$}).
\end{equation}
In this equation we are representing the Pauli spin matrices as a
vector, $\mbox{\boldmath $\sigma$} =  {\bf \hat{i}\sigma_x +
\hat{j}\sigma_y + \hat{k}\sigma_z}$. Thus $\mbox{\boldmath
$\sigma\cdot a$} = \sigma_xa_x + \sigma_ya_y + \sigma_za_z$
represents a traceless, $2 \times 2$ matrix and {\bf 1} is the
unit matrix. Now, the expectation value of any operator ${\cal O}$
can be written $<{\cal O}> =\rm Tr(\rho {\cal O})$, where $\rho$
is the density matrix $\equiv \rm diag (1/2,1/2)$ for an initially
unpolarized beam. The probability of finding the first particle in
the + state along {\bf a} is thus $\rm Tr(\rho\Pi({\bf a})) =
1/2$. Similarly, the joint probability $P({\bf a}+,\ {\bf b}\pm)$
of finding the first particle in the + state along {\bf a} and the
second particle in the $\pm$ state along {\bf b} is
\begin{eqnarray}
P({\bf a}+, {\bf b}\pm) & = & \frac{1}{2}Tr \Pi({\bf a})\Pi({\bf
                                            b}\pm) \nonumber \\
 & = & \frac{1}{8} Tr \{({\bf 1} + \mbox{\boldmath $\sigma\cdot a$})
                    ({\bf 1} \pm \mbox{\boldmath $\sigma\cdot b$})\}
                                                    \nonumber \\
 & = & \frac{1}{4} ( 1 \pm {\bf a \cdot b}).
\label{Malus}
\end{eqnarray}
Here, use has been made of the standard identity (see
\cite{Sakurai})
\begin{equation}
(\mbox{\boldmath $\sigma\cdot a$})(\mbox{\boldmath $\sigma\cdot
b$}) = {\bf (a\cdot b)1} + i\mbox{\boldmath$\sigma$} {\bf (a
\times b)}. \label{standard}
\end{equation}
Because the Pauli matrix is traceless, taking the trace of
(\ref{standard}) yields 2{\boldmath $a\cdot b$}.

Equation (\ref{Malus}) is simply a sophisticated way of writing
Malus' law. The first factor of $1/2$ in (\ref{Malus}) gives the
probability of detecting a particle in the + state along the {\bf
a} axis. The remaining factor  $1/2(1 + \mbox{\boldmath $a\cdot
b$})= 1/2 (1 + cos\theta)$, where $\theta$ is the angle between
polarizers. For photons(where $\theta$ is taken to be the double
angle) this then represents the usual decrease in intensity with
$cos^2\theta$. For a Bohm-type experiment,which assumes an
(antisymmetric) spin-singlet state, one should choose the $-$ on
the right of (\ref{Malus})when computing $P({\bf a}+, \ {\bf b}
+)$ to conserve angular momentum. With either sign, by inserting
(\ref{Malus}) into (\ref{Bell}), it is straightforward to show
that quantum mechanics violates Bell's inequalities.

For the Eberhard experiment we take the knob settings
$a_1,a_2,b_1,b_2$  to represent the position of the polarizers on
the measuring devices.  Recall that his quantities $C =
<\alpha\beta>$ represented the fraction of events in which
$\alpha$ and $\beta$ had the same sign minus the fraction in which
they had opposite signs, irrespective of whether an individual
spin is $+$ or $-$. Evidently the equivalent quantum expression is
$1/2(1 + {\bf a\cdot b})-1/2(1 - {\bf a\cdot b})$. Then
\begin{equation}
{\cal C}_{qm} = {\bf a_1\cdot b_1}
                + {\bf a_1\cdot b_2}
                + {\bf a_2\cdot b_1}
                - {\bf a_2\cdot b_2}.
                \label{Cqmb}
\end{equation}
If the axes are chosen to be coplanar such that ${\bf a_1\cdot
b_1} = {\bf a_1\cdot b_2} = {\bf a_2\cdot b_1}= cos\theta$ and
${\bf a_2\cdot b_2}= cos3\theta$, then (\ref{Cqmb}) gives exactly
(\ref{Cqm}), which violates the CHSH inequality for $\theta =
45^o$.

The derivation of (\ref{Cqm}) just given involved only pairwise
probabilities and did not go beyond standard quantum mechanics.
With the projection-operator formalism, however, it is not
difficult to write down the joint probability for four
``simultaneous" spin measurements among four axes. An example
would be $P(++++)$, in analogy to the classical three-probability
mentioned earlier that appears in the derivation of Bell's
inequality. Extending (\ref{Malus}) to four arguments we take
\begin{equation}
P({\bf \lambda a_1, \mu a_2, \nu b_1, \tau b_2}) =
        \frac{1}{2}Tr \{\Pi(\lambda {\bf a_1})\Pi(\mu {\bf a_2})\Pi(\nu
        {\bf b_1})\Pi(\tau {\bf b_2})\},
\end{equation}
where $\lambda,\mu,\nu,\tau$ are chosen as $\pm 1$ to represent up
or down.  For the symmetric case this is
\begin{equation}
P({\bf \lambda a_1, \mu a_2, \nu b_1, \tau b_2})=
    \frac{1}{32}Tr \{
                    ({\bf 1} + \lambda\mbox{\boldmath $\sigma\cdot a_1$})
                    ({\bf 1} + \mu\mbox{\boldmath $\sigma\cdot a_2$})
                    ({\bf 1} + \nu\mbox{\boldmath $\sigma\cdot b_1$})
                    ({\bf 1} + \tau\mbox{\boldmath $\sigma\cdot b_2$})
                      \}
\label{P4}
\end{equation}
We will need the antisymmetric expression later to make the
subtraction just done above. Assuming that a measurement of + on
knob $a$ requires  $-$ on  knob $b$, the antisymmetric case will
be the same expression as (\ref{P4})with the signs on the $b$'s
reversed. We calculate only the symmetric case and state the
results for the antisymmetric case as needed.

Working out (\ref{P4}) and making frequent use of the identity
(\ref{standard}) yields
\begin{eqnarray}
P({\bf \lambda a_1, \mu a_2, \nu b_1, \tau b_2})=
                  \frac{1}{16} \{ 1 +
                  {\bf \lambda\mu a_1\cdot a_2 + \lambda\nu
                  a_1\cdot b_1 + \lambda\tau a_1\cdot b_2 } \nonumber \\
                  {\bf + \mu\nu
                  a_2\cdot b_1 + \mu\tau a_2\cdot b_2 + \nu\tau
                  b_1\cdot b_2 }\nonumber \\
                 {\bf  + \imath\lambda\mu\nu (a_1 \times a_2)\cdot b_1
                  + \imath\lambda\mu\tau (a_1 \times a_2)\cdot b_2 }\nonumber \\
                 {\bf + \imath\lambda\nu\tau (b_1 \times b_2)\cdot a_1
                  + \imath\mu\nu\tau (b_1 \times b_2)\cdot a_2 }\nonumber \\
                  {\bf +\lambda\mu\nu\tau
                  \left[(a_1\cdot a_2)(b_1\cdot b_2)
                  +2(a_1 \times a_2)\cdot(b_1 \times b_2)\right]
                  }\}.
                  \label{P4b}
\end{eqnarray}
Notice that this expression is complex due to the imaginary
elements of $\sigma_y$.  If we desire a real result to eventually
make contact with the usual quantum predictions, we can easily
eliminate the imaginary terms.  Note that $\Pi(\lambda {\bf
a_1})\Pi(\mu {\bf a_2})\Pi(\nu {\bf b_1})\Pi(\tau {\bf b_2})$ has
been written in an arbitrary order; it is not symmetric in the
arguments.  There are $4!$ permutations of the arguments in this
expression, twelve even and twelve odd. In (\ref{P4b}) each
imaginary term is a triple scalar product, which is invariant
under even permutations and changes sign under odd permutations.
Thus these terms vanish under symmetrization, as does the double
cross product in the last line. The symmetrized version of
(\ref{P4b}) is
\begin{eqnarray}
P({\bf \lambda a_1, \mu a_2, \nu b_1, \tau b_2})=
                  \frac{1}{16} \{ 1 +
                  {\bf \lambda\mu a_1\cdot a_2 + \lambda\nu
                  a_1\cdot b_1 + \lambda\tau a_1\cdot b_2 } \nonumber \\
                  {\bf + \mu\nu
                  a_2\cdot b_1 + \mu\tau a_2\cdot b_2 + \nu\tau
                  b_1\cdot b_2 }\nonumber \\
                  + \frac{1}{3} {\bf  \lambda\mu\nu\tau
                  \left[(a_1\cdot a_2)(b_1\cdot b_2)
                  +(a_1\cdot b_1)(a_2\cdot b_2)
                   +(a_1\cdot b_2)(b_1\cdot a_2) \right]
                  }\},
                  \label{P4s}
\end{eqnarray}
which is entirely real. \footnote{It is not actually necessary to
symmetrize (\ref{P4b}).  One can leave it as a complex expression,
but when the sum over the extraneous arguments is performed as in
(\ref{Sum}), the imaginary terms cancel and the result will be
entirely real, as before. However, the complex four-probability is
not symmetric in the arguments.}

It is now easy to read off the various four-probabilities,
$P(++++), P(- - - -)$ etc. for each case merely by choosing the
required signs of $\lambda,\mu,\nu,\tau$.  The sixteen
possibilities are listed for convenience in Table II. Note that
these four-probabilities do sum to one and therefore in that
respect behave as ordinary probabilities. However, although it is
perhaps not evident from inspection, several of these
probabilities can become negative. We plot $P(+++-)$ and $P(+-+-)$
in Figure 1. The antisymmetric $P$'s can be obtained from the
symmetric ones merely merely by flipping the signs on the two
$b$'s.

From these four-probabilities one can form the quantity ${\cal
C}_{qm}$ in Eq. (\ref{Cqmb}) in exact analogy to the procedure
used for deriving the Bell inequalities.  To compute $P({\bf a_1+,
b_1+})$, for example, we only care that the first particle will be
found + along $\bf a_1$ and the second particle will be found $+$
along $\bf b_1$.  As before, we count all such possibilities by
summing over the two extraneous arguments, $\bf a_2$ and $\bf
b_2$. Thus, for the symmetric wavefunction,
\begin{equation}
P({\bf a_1+, b_1+}) = P(+ \_\_ + \_\_ ) = P(++++) +
                                    P(+++-)+P(+-+-)+P(+-++)
\label{Sum}
\end{equation}
 Reading off these $P$'s from Table II and performing the sum
yields
\begin{equation}
\frac{1}{4}(1+ {\bf a_1\cdot b_1}),
\end{equation}
which is exactly Eq. (\ref{Malus}).  For the antisymmetric wave
function one obtains $1/4(1 - {\bf a_1\cdot b_1})$. Similar
expressions are obtained for the other three pairwise
probabilities.  Clearly, subtracting the antisymmetric expressions
from the symmetric ones and adding the four terms leads back to
Eq. (\ref{Cqmb}) for ${\cal C}_{qm}$. This procedure must work
because the four-probabilities are symmetric in all the arguments;
summing over any of them produces an equal number of terms of
opposite sign, which cancel out, leaving the usual quantum
pairwise probabilities.

\section{Discussion and Conclusions}
\setcounter{equation}{0}
 \label{Conclusions}

We have shown that, like the Bell inequalities, the CHSH
inequality assumes positive-definite probabilities and that
quantum mechanics breaks both inequalities effectively because it
introduces negative weights to the measurements.  These negative
four-probabilities enter the derivation in exactly the same way as
the classical three-probabilities entered the derivation of the
Bell's inequalities.  If they are unphysical, it is not
necessarily because they are negative, but because it is
impossible to make four simultaneous spin measurements on two
particles.  By the same token, it is impossible to make three
simultaneous spin measurements on two particles.  In any case,
neither the classical three-probabilities found in Bell's theorem,
nor the four-probabilities that figure here are actually measured.
Both merely serve as ``master distributions" from which to derive
the usual pairwise probabilities, classical and quantum, which are
both positive-definite. To reiterate our earlier remarks, from
this point of view it is not surprising that the Bell and CHSH
inequalities are violated by experimental tests; they merely used
the wrong set of probabilities for a quantum-mechanical problem.

Although one might choose to reject negative probabilities as
unphysical, one should not reject the notion of master probability
distributions in favor of correlations between real and imaginary
experiments because the two procedures are identical!  Recall
again that Eberhard's quantity $C_{11}$ was $C_{11} =
\frac{1}{N}\sum_{j=1}^N\alpha_{1j}\beta_{1j}$, which represented
the fraction of events $\alpha_1\beta_1$ that had the same sign
minus the fraction that had opposite sign.  Thus by definition we
can write
\begin{equation}
C_{11} = P({\bf a_1+, b_1+}) + P({\bf a_1-, b_1-})
             -[P({\bf a_1+, b_1-})+P({\bf a_1+, b_1-})].
\end{equation}
Now, in exact analogy with the procedure of Section \ref{QMP} we
imagine that these pairwise probabilities can be derived from a
master distribution involving all four axes ${\bf a_1, a_2, b_2,
b_3}$.  In that case, as in Eq (\ref{Sum}), $P(++)= P({\bf a_1+,
b_1+}) = P(++++) + P(+++-)+P(+-+-)+P(++-+)$, with analogous
expressions for $P(--), P(+-)$ and $P(-+)$.  There are thus 16
terms that contribute to $C_{11}$, similarly for $C_{12}, C_{21}$
and $C_{22}$. Writing out all 64 terms yields for ${\cal C} =
<\gamma>$:
\begin{eqnarray}
{\cal C} = &&
          2\{
            P(++++) + P(----) +P(+++-) + P(---+)\nonumber\\
         && + P(+-++) + P(-+--) + P(+--+) + P(-++-)\nonumber\\
         && - P(++-+) - P(--+-) - P(-+++) -P(+---)\nonumber\\
         && - P(++--) - P(--++) - P(+-+-) - P(-+-+)
         \}
\label{CP}
\end{eqnarray}
These $P$'s are general and may be taken to be either classical or
quantum. Notice half enter with positive sign and half with
negative. If all the probabilities are equal, then ${\cal C} = 0$.
If those that enter with negative sign are zero, then ${\cal C} =
2$ and if those that enter with positive sign are zero, then
${\cal C} = -2$. All this is in complete agreement with the
analysis of Section \ref{sec2}. Clearly, if the $P$'s are
positive-definite then ${\cal C} \le 2$, but if the probabilities
are allowed to become negative then this bound is violated. If the
$P$'s are assumed to be quantum, they take on the values given by
Table II. In this case, inserting those values into (\ref{CP})
gives exactly (\ref{Cqmb}), as before.

This demonstration shows clearly that the $\gamma$'s can be
derived from a master probability distribution which involves
simultaneous spin measurements along four axes. The {\em only}
difference between the classical and quantum cases is that in the
former we assume the probabilities are positive-definite. The
master distributions themselves cannot be regarded as any more or
less meaningful than the space of hypothetical measurements,
because the procedures are exactly equivalent.  Indeed, we see
that there is no difference between the Eberhard procedure and the
usual derivation of Bell's inequalities.

There remains the problem of interpretation.  Most people insist
that probability be defined in terms of relative frequency of
events, in which case it must be positive-definite. In quantum
mechanics, however, although one can define the expectation value
in terms of the square of the wave amplitude, which corresponds to
a relative-frequency interpretation, an alternate procedure is
available. The expectation value may also be taken as a functional
of the dynamical variables under consideration, for example
position {\it and} momentum. Classically, one might consider a
Maxwellian distribution of particles in phase space; integrating
over position or momentum would give the marginal probability
distribution for the conjugate variable. But in quantum mechanics,
the uncertainty principle precludes precise simultaneous knowledge
of noncommuting variables. If one attempts to associate a
functional with a distribution over noncommuting variables, such
that an integration over one of them gives the correct marginal
distribution for the other, then one finds that the distribution
function must in places become negative.  This is the well known
Wigner Distribution\cite{Wigner32}.

In the case of spin, the different components of angular momentum
do not commute; hence no ordinary (positive-definite) probability
distribution can be defined over the various components
simultaneously.  Any distribution will share with the Wigner
distribution the property that it will become negative in some
region of ``phase space."   For example, in the spin-1/2 systems
we have been considering, the probability of finding $S_z$ in the
$+$ state and $S_x$ in the $+$ state is given by taking the trace
of the product of the projection operators, as we have done
earlier.  Now, given a state with $S_x = +$, the probability is
1/2 for finding $S_z = +$, and 1/2 for $S_z = -$. Suppose,
however, that many measurements show $S_z = +$, always, but that
$S_x = +$ appears with probability $\lambda$ and $S_x = -$ appears
with probability $1 - \lambda $ ($ 0 \le \lambda \le 1$).  The
probability for finding $S_z = -$ must be then be $(1/2)\lambda +
(1/2)(1-\lambda) = 1/2$. On the one hand the probability of $S_z =
-$ must equal zero.  On the other hand, no mixture of $S_x = +$
and $S_x = -$ can give a zero probability for $S_z = -$.

This is quite a general property of noncommuting variables and has
little to do with quantum mechanics.  In such situations the best
that one can ask for is that the probability distribution give the
correct marginal distribution for one of the variables, in our
case one component of angular momentum.  This is what has been
found in the present paper. The probability distribution for
simultaneous measurements along three or more axes are not
positive-definite, but the marginal distributions that give
correlations between two spin components are, and are in accord
with the standard predictions of quantum mechanics.

The main point of this paper has been that assumptions beyond
locality do enter into derivations of Bell's inequalities. It is
worth mentioning yet another tacit assumption: that space is flat.
The notion of parallel and antiparallel spins is only well defined
for flat space where the measurement axes (the ``z" axes) can be
taken to be everywhere fixed relative to one another. In curved
space there is no universal definition of parallel and one can
only compare spins in distant locations by parallel transporting
the measurement axes \cite{BM00}.  In the case of nonnegligible
gravitational fields, then, the ``nonlocal" EPR correlation
between two particles, to the extent that they can be said to
exist at all, must be the result of parallel transport, a local
phenomenon.

Returning to probabilities, we find ourselves in a strange
situation. If one insists that probabilities remain
positive-definite, we are forced to use vague and imprecise
concepts, such as ``local" or ``nonlocal" to describe the outcome
of the EPR experiment.  On the other hand, we are able formulate
the precise mathematical conditions necessary for the violation of
the Bell and CHSH inequalities, although at the cost of
introducing negative probabilities.  Most investigators would say
that a unified, physical interpretation of negative probabilities
is, in fact, exactly what is currently lacking. To be sure,
Feynman conceded (see \cite{Muck86} and \cite{Feynman48}; also
\cite{Sud63,Mehta65})that all the results of quantum mechanics can
be analyzed in terms of negative probabilities but he remained
skeptical about the utility of such an approach and that a useful
meaning could be attached to it. Nevertheless, many of the
interpretational problems associated with negative probabilities
stem from an insistence on viewing them within the framework of
relative frequencies. This is clearly ``no go." We have shown that
a more natural framework for their interpretation arises when one
considers the expectation value as a measure of probability over
noncommuting variables.  One can even go further than we have and
consider complex probability measures (\cite{Srin94}), which also
involve expectation values. Under such circumstances it is well to
bear in mind that imaginary numbers are more similar to rotations
than to real numbers.  One should also bear in mind the very word
``imaginary," an obsolete relic of their original status.\\

\noindent{\bf Acknowledgements} We would like to thank Sebastiano
Sonego for bringing our attention to the Eberhard and Peres papers
and explaining a few details of the former. T.R. would also like
to thank Gabe Spalding for helping to check some algebra.\\

\noindent{\bf Note added:} Since this paper was initially posted,
Jos\'e Cereceda has come to essentially the same conclusions (see
quant-ph/0010091).

\section*{Appendix}
\setcounter{equation}{0}
 \label{Appendix}
\renewcommand{\theequation}{A.\arabic{equation}}

Many researchers appear unwilling to accept that {\em any} assumptions beyond
locality are employed in the derivations of Bell's inequalities.  We now list
a few of the proofs we have found and point out explicitly where the
assumption of positive probabilities enters.

{\it Bell 64}.  In Bell's original proof \cite{Bell64} he defines
two quantities $A(\vec{a},\lambda) = \pm 1$, $B (\vec{b},\lambda)
= \pm 1$.  He defines a normalized probability distribution
$\rho(\lambda)$, such that $\int d\lambda \;\rho(\lambda) = 1$.
The expectation value of the spin components $\vec{\sigma_1}\cdot
\vec{a}$ and $\vec{\sigma_2}\cdot \vec{b}$ is
\begin{equation}
P(\vec{a},\vec{b}) = \int d\lambda \;\rho(\lambda)
                A(\vec{a},\lambda)B (\vec{b},\lambda),
\end{equation}
which he shows can be written (his equation 14) as
\begin{equation}
P(\vec{a},\vec{b}) = - \int d\lambda \;\rho(\lambda)
                A(\vec{a},\lambda) A(\vec{b},\lambda).
\end{equation}
When another vector $\vec{c}$ is involved, one has
\begin{equation}
P(\vec{a},\vec{b}) - P(\vec{a},\vec{c})=
         - \int d\lambda \;\rho(\lambda) [
         A(\vec{a},\lambda)A (\vec{b},\lambda) -
         A(\vec{a},\lambda)A (\vec{c},\lambda)]
\end{equation}
Bearing in mind that$A(\vec{b},\lambda) = 1/A(\vec{b},\lambda)$ one can rewrite
this as
\begin{equation}
P(\vec{a},\vec{b}) - P(\vec{a},\vec{c})=
         \int d\lambda \;\rho(\lambda)
         A(\vec{a},\lambda)A(\vec{b},\lambda)
             [A(\vec{b},\lambda)A (\vec{c},\lambda) -1].
\end{equation}
Bell then asserts
\begin{equation}
|P(\vec{a},\vec{b}) - P(\vec{a},\vec{c})| \le
         \int d\lambda \;\rho(\lambda)
             [A(\vec{b},\lambda)A (\vec{c},\lambda) -1],
\label{BellA}
\end{equation}
where, of course, $|A(\vec{a},\lambda)A(\vec{b},\lambda)| = 1$.  However,
stricly speaking the triangle inequality gives
\begin{equation}
|P(\vec{a},\vec{b}) - P(\vec{a},\vec{c}) | \le
         \int d\lambda \;|\rho(\lambda)|
            [A(\vec{b},\lambda)A (\vec{c},\lambda) -1],
\end{equation}
which is equal to (\ref{BellA}) only when $|\rho| = \rho$, i.e.,
when $\rho \ge 0$.

{\it CHSH}.  The CHSH paper \cite{CHSH69} makes the same
assumption at the identical point in their derivation, in their
first (unnumbered) equation.

{\it Peres}.  Peres' derivation \cite{Peres78} is almost identical to Eberhard's
and makes the same assumption of positive weights in the same step, i.e.
between steps 1 and 2 of Eq. (\ref{triangle}) of this paper.

{\it Stapp 71}.  Stapp's 1971 proof \cite{Stapp71} is very similar
to Bell's. He arrives at an expression (below his equation 8)
\begin{equation}
\sqrt 2 \le \frac{1}{N}\sum_j|n_{2j}''n_{2j}' - 1|,
\end{equation}
where $n_{2j}'' = \pm 1$ and $n_{2j}' = \pm 1$.  He then shows
this leads to the contradiction $\sqrt 2 \le 1$.  However, if the
$n$'s are $\pm 1$, then the summand can only have values 0,2. If
$N_1$ and $N_2$ are the frequencies with which these two values
occur, and $N_1 + N_2 = N$, then the right hand side can be
written
\begin{equation}
\frac{1}{N}[N_1 \times 0 + N_2 \times 2] = \frac{2N_2}{N} =
\frac{2(N - N_1)}{N} = 2(1 - \frac{N_1}{N}).
\end{equation}
As in the Eberhard argument, a contradiction can always be avoided
by taking $N_1$ negative.

{\it Stapp 85}.  Stapp \cite{Stapp85} establishes a contradiction by
demonstrating (his Eq. 8) that
\begin{equation}
\frac{1}{n}\sum_{i = 1}^n\left[\sqrt{2}\; r_{Ai}(\hat\lambda_a) +
            r_{Bi}(\hat\lambda_a) + r_{Bi}(\hat\lambda_b)\right]^2
        > (\sqrt 2 - 2)^2,
\end{equation}
where $r_{Ai}(\hat\lambda_a) = \pm 1$, $ r_{Bi}(\hat\lambda_a) =
\pm 1$ and $r_{Bi}(\hat\lambda_b) = \pm 1$ . However, since the
$r$'s are $\pm 1$, the summand can have only one of three values:
$(\sqrt 2)^2$, $(2 + \sqrt 2)^2$ and $(2 -\sqrt 2)^2$. Then the
above expression can be written as
\begin{equation}
\frac{1}{n}\left[n_1(\sqrt{2})^2 + n_2 (\sqrt 2 + 2)^2 + n_3(2 - \sqrt 2)^2
            \right],
\end{equation}
where $n_1, n_2, n_3$ are the frequencies with which the three terms occur
and $n_1 + n_2 + n_3 = n$.  Squaring out and combining terms yields
\begin{equation}
\frac{2(n_1 + n_2 + n_3)}{n} + \frac{2n_2(2 + \sqrt2)}{n}
        + \frac{2n_3(2-\sqrt 2)}{n}.
\end{equation}
Assuming $n$ and $n_3$ positive, this expression can become
negative if
\begin{equation}
n_2 < \frac{-(n + n_3(2 - \sqrt 2))}{2 + \sqrt 2},
\end{equation}
in other words, if $n_2$ is sufficiently negative.

{\it Bell 71}.  A proof that has been cited as qualitatively
different than the others is Bell's 1971 proof {\cite{Bell71}.
This proof is basically the same as the CHSH proof. In Bell's 1971
version the probability density is also explicitly taken to be
positive definite. The only difference is that now
$|A(\vec{a},\lambda)| \le 1$ and $|B (\vec{b},\lambda)| \le 1$.
(In our notation this corresponds to $|\alpha_i| \le 1$ and
$|\beta_i| \le 1$.) This change merely strengthens the upper bound
on the classical correlations. That is, in our equation
(\ref{gambound}), whereas previously $|\gamma| = 2$, now $|\gamma|
\le 2$.  The rest of the derivation is consequently unaffected and
the CHSH inequality continues to hold. Furthermore, our
demonstration of the equivalence of the Eberhard procedure with
the "master probability distribution" procedure is also
unaffected, since Eq. (\ref{CP}) made no assumption about the
values of the $P$'s.

\newpage

\begin{center}
{\bf TABLES AND FIGURES}
\end{center}

TABLE I. Spin combinations for standard Bell inequalities.
Hidden-variable models assume that spin-1/2 particles can be
emitted with $\pm$ spin along each of three axes, {\bf a}, {\bf b}
and {\bf c}. The notation ($+++$) etc., means spin up along all
three axes.  The eight possible spin combinations are shown. To
ensure conservation of angular momentum, a particle of the type
$(+++)$ must be paired with one of ($---$) and so on.
\begin{center}
 \vspace{1in}
\begin{tabular}{ccc}
Population  &       Particle 1  &   Particle 2 \\ \hline\hline

 $ N_1$     &       ($+++$)     &   ($---$)  \\
 $N_2$      &       ($++-$)     &   ($--+$) \\
 $N_3$      &       ($+-+$)     &   ($-+-$) \\
 $N_4$      &       ($-++$)     &   ($+--$) \\
 $N_5$      &       ($+--$)     &   ($-++$) \\
 $N_6$      &       ($-+-$)     &   ($+-+$) \\
 $N_7$      &       ($--+$)     &   ($++-$) \\
 $N_8$      &       ($---$)     &   ($+++$) \\\hline\hline
 \end{tabular}
\end{center}

\newpage

TABLE II. Four probabilities.  Shown are the four-probabilities
from symmetric wavefunction as computed from Eq. (\ref{P4s}). The
quantity
 \begin{center} $\Delta \equiv \frac{1}{3} {\bf
\left[(a_1\cdot a_2)(b_1\cdot b_2) +(a_1\cdot b_1)(a_2\cdot b_2)
+(a_1\cdot b_2)(b_1\cdot a_2) \right]}$.
\end{center}
 Note that these probabilities sum
to one. The four-probabilities for the antisymmetric wave function
can be obtained by flipping last two signs, i.e., $P(++++)_{AS} =
P(++--)_s, \ P(-+++)_{AS} = P(-+--)_s $, etc.
\begin{center}
\begin{tabular}{c}
\vspace{5mm}\\ \hline \\  \vspace{2mm}

 $P(++++) = P(- - - -) = \frac{1}{16} \{1 + {\bf a_1 \cdot a_2 +
a_1 \cdot b_1 + a_1 \cdot b_2 + a_2 \cdot b_1 + a_2 \cdot b_2 +
b_1 \cdot b_2 + }\Delta \} $

 \vspace{1mm} \\ \hline
\\  \vspace{1mm}
$ P(-+++) = P(+ - - -) = \frac{1}{16} \{1 - {\bf a_1 \cdot a_2 -
a_1 \cdot b_1 - a_1 \cdot b_2 + a_2 \cdot b_1 + a_2 \cdot b_2 +
b_1 \cdot b_2 - }\Delta \} $

\vspace{1mm} \\  \hline
\\  \vspace{1mm}

$P(+-++) = P(- + - -) = \frac{1}{16} \{1 - {\bf a_1 \cdot a_2 +
a_1 \cdot b_1 + a_1 \cdot b_2 - a_2 \cdot b_1 - a_2 \cdot b_2 +
b_1 \cdot b_2 - }\Delta \}$

\vspace{1mm} \\  \hline
\\  \vspace{1mm}
$
P(++-+) = P(- - + -) = \frac{1}{16} \{1 + {\bf a_1 \cdot a_2 - a_1
\cdot b_1 + a_1 \cdot b_2 - a_2 \cdot b_1 + a_2 \cdot b_2 - b_1
\cdot b_2 - }\Delta \}$

 \vspace{1mm} \\  \hline
\\  \vspace{1mm}

$ P(+++-) = P(- - - +) = \frac{1}{16} \{1 + {\bf a_1 \cdot a_2 +
a_1 \cdot b_1 - a_1 \cdot b_2 + a_2 \cdot b_1 - a_2 \cdot b_2 -b_1
\cdot b_2 - }\Delta \}$

 \vspace{1mm} \\  \hline
\\ \vspace{1mm}

$ P(++--) = P(- - + +) = \frac{1}{16} \{1 + {\bf a_1 \cdot a_2 -
a_1 \cdot b_1 - a_1 \cdot b_2 - a_2 \cdot b_1 - a_2 \cdot b_2 +
b_1 \cdot b_2 + }\Delta \}$

 \vspace{1mm} \\  \hline
\\  \vspace{1mm}

$ P(+-+-) = P(- + - +) = \frac{1}{16} \{1 - {\bf a_1 \cdot a_2 +
a_1 \cdot b_1 - a_1 \cdot b_2 - a_2 \cdot b_1 + a_2 \cdot b_2 -
b_1 \cdot b_2 +  }\Delta\}$

 \vspace{1mm} \\  \hline
\\  \vspace{1mm}
$
 P(+ - - +) = P(- + + -) = \frac{1}{16} \{1 - {\bf a_1 \cdot a_2 -
a_1 \cdot b_1 + a_1 \cdot b_2 + a_2 \cdot b_1 - a_2 \cdot b_2 -
b_1 \cdot b_2 +  }\Delta\}$

\vspace{1mm} \\  \hline
\end{tabular}
\end{center}

\newpage

TABLE III. Four probabilities as functions of polarizer angles.
Shown are the same four-probabilities as on Table II for the
configuration ${\bf a_1 \cdot b_1= a_1 \cdot b_2 = b_1 \cdot a_2}
= cos\theta$ and  ${\bf a_2 \cdot b_2} = cos3\theta $. Now $\Delta
= 1/3(cos^2\theta + cos^22\theta + cos\theta\cos3\theta) $.  With
the identities $cos2\theta = 2cos^2\theta -1$ and $cos3\theta = 4
cos^3\theta-3cos\theta $ all the probabilities can be written in
terms of one parameter, $cos\theta \equiv C$.  This form makes it
more plausible that some of the $P$'s can become negative.
\begin{center}
\begin{tabular}{l}
\vspace{5mm}\\ \hline \\  \vspace{2mm}

 $P(++++) = P(- - - -) = \frac{1}{16} \{1 + 3cos\theta
                         + 2cos2\theta + cos3\theta + \Delta\}
                       = \frac{1}{16} \{4C^3+4C^2 - 1 + \Delta
                       \}$

 \vspace{1mm} \\ \hline
\\  \vspace{1mm}
$ P(-+++) = P(+ - - -) = \frac{1}{16} \{1 -cos\theta
                        + cos3\theta-\Delta \}
                      = \frac{1}{16} \{4C^3- 4C + 1 -\Delta \}
$

\vspace{1mm} \\  \hline
\\  \vspace{1mm}

$P(+-++) = P(- + - -) = \frac{1}{16} \{1 + cos\theta
                         - cos3\theta - \Delta \}
                         = \frac{1}{16}\{-4C^3 + 4C + 1 - \Delta\} $

\vspace{1mm} \\  \hline
\\  \vspace{1mm}
$
P(++-+) = P(- - + -) = \frac{1}{16} \{1 - cos\theta
                        + cos3\theta - \Delta \}
                     =  \frac{1}{16} \{4C^3- 4C + 1 -\Delta \} $
 \vspace{1mm} \\  \hline
\\  \vspace{1mm}

$ P(+++-) = P(- - - +) = \frac{1}{16} \{1 + cos\theta
                        - cos3\theta - \Delta \}
                         =  \frac{1}{16}\{-4C^3 + 4C + 1 - \Delta\} $
 \vspace{1mm} \\  \hline
\\ \vspace{1mm}

$ P(++--) = P(- - + +) = \frac{1}{16} \{1 + 2 cos2\theta
                            -3cos\theta - cos3\theta + \Delta\}
                        =  \frac{1}{16} \{-4C^3 + C^2 - 1 +
                        \Delta\}$

 \vspace{1mm} \\  \hline
\\  \vspace{1mm}

$ P(+-+-) = P(- + - +) = \frac{1}{16} \{1 - cos\theta
                         - 2cos2\theta + cos3\theta + \Delta\}
                        =  \frac{1}{16} \{4C^3 - 4C^2 - 4C + 3 +
                        \Delta\}$
 \vspace{1mm} \\  \hline
\\  \vspace{1mm}
$
 P(+ - - +) = P(- + + -)  = \frac{1}{16} \{1 + cos\theta
                         - 2cos2\theta - cos3\theta + \Delta\}
                        =  \frac{1}{16} \{-4C^3 - 4C^2 + 4C + 3 +
                        \Delta\}$

\vspace{1mm} \\  \hline
\end{tabular}
\end{center}

\newpage
FIG. 1.  Four-probabilities from Table III. (a) Plot of
$16P(+++-)$. (b) Plot of $16P(+-+-)$. Note that these quantities
become negative.
\begin{figure}
\includegraphics{qm_fig1a.ps}
\end{figure}
\begin{figure}
\includegraphics{qm_fig1b.ps}
\end{figure}
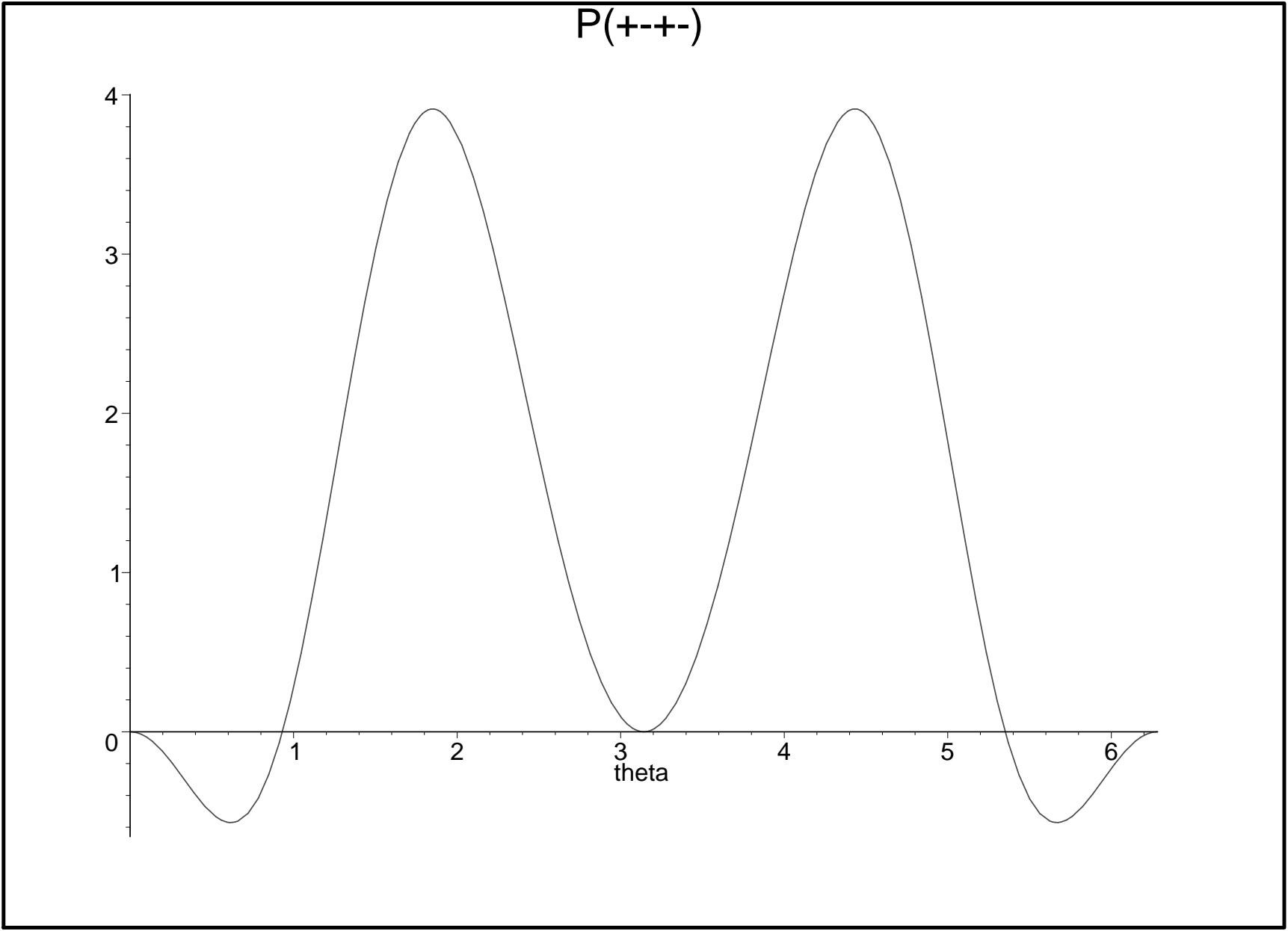
\end{document}